\documentstyle{article}
\begin{document}

\begin{flushright}
ITP-NENU\hspace{0.5cm} 94-11
\end{flushright}

\Large
\begin{center}
{\bf Quantum Dynamics for the Control of Atomic State by a
Quantized Optical Ring Cavity}
\vspace{2cm}
\Large

{\it Chang-Pu Sun\\ }
Institute of Theoretical Physics, Northeast Normal University, Changchun
130024, P.R.China

Abstract
\end{center}
\large
A generalized approach of the Born-Oppenheimer approximation is developed
to analytically deal with the influence exercised by the spatial motion
of atom's mass-center on a two-level atom in an optical ring cavity
with a quantized single-mode electromagnetic field. The explicit expressions
of tunneling rate are obtained for various cases, such as that with initial
coherent state and thermal equilibrium state at finite temperature. Therefore,
the studies for Doppler and recoil effects of the spatial motion
on the scheme controlling atomic tunneling should be reconsidered in terms of
the initial momentum of  atom's mass center.
\newpage
{\bf 1.Introduction}
\vspace{0.5cm}

Many new developments in both the experimental and theoretical aspects of the
so-called cavity quantum electrodynamics [1,2] have shown the possibility
controlling the coherent tunneling of atomic states by a quantized or
classical cavity field [3-7]. By immersing an atom into a cavity
field with a proper strength and frequency, the tunneling rate can
be enhanced or reduced to the values several
orders of magnitude higher than the rate for the ``bare" atom.

To understand the essence in the mechanism governing the control of atomic
tunneling, we draw
an obvious analogy in the basic quantum mechanics, the coherent tunneling
phenomenon in a one-dimensional double well potential. It is well-known that
the two lowest-energy states with a finite split of energy
in this potential well
possess even and odd parities respectively. Their symmetric and antisymmetric
superpositions approximately represent the localizations of particle
in the right and left wells separately. Due to the energy split, one of the
two superposition states can evolve into another and then back to the
original one. Such
a coherent tunneling with a period determined by the energy split enjoys
the general feature of quantum mechanics, but the phenomenon of
localization does not appear in the usual case that the system is
isolated as a closed system.
To change the tunneling rate, a possible way is to immerse the system
in a certain environment as an open system, but the control for a given
goal can not be realized in this sense because of the random elements such
as the Brownian motion exercised by the environment [8]. However, in the
above-mentioned studies about the control of atomic tunneling
, the random environment is replaced by the cavity field, which can
be prepared in advance for a specific purpose to enhance or  reduce
the tunneling rate. Notice that all the investigations about atomic
tunneling control have not concerned the motion of mass-center to our
best knowledge.

In this paper, based on the generalized Born-Oppenheimer (BO) approximation
developed by this author to separate the fast and slow dynamical
variables for the spin-precession in an inhomogeneous magnetic field [9, 10],
we present a delicate study  to analyse
the influence of the motion of atomic mass-center on the tunneling and
localization of atomic states in a cavity field. In fact, the spatial
motion of atom (strictly speaking, its mass-center) plays a crucial role in
many fashionable problems in so-called atom-optics, such as the diffraction
and  splitter of atom beam by a standing wave cavity field in connection
with atom interferometer [11-15], the quantum nondemolition measurement by
an optical ring cavity [16] and trapping and colling atoms with
an adiabatically-decaying cavity mode [17,18]. In the limit of  strong field,
the dressed atomic eigenstates are obtained in accord with the generalized
BO approximation. It is then proved that the higher order approximations
mix them to cause the tunneling from one to another among them. The
explicit expressions of tunneling rates are given to manifest the crucial
role of the Doppler effect of spatial motion of atomic mass center
in a locally-inhomogeneous cavity field.
\vspace{1cm}

{\bf 2.The model }
\vspace{0.5cm}

Consider the most simple case that a two-level atom moves along the optical
axis $x$ in an optical ring cavity with a single-mode quantized electromagnetic
field
\begin{equation}
E\sim a^{\dagger} e^{-ik\hat{x}} + a e^{ik\hat{x}}
\end{equation}
where $a^{\dagger}$ and $a$ are the creation and annihilation operators
for the cavity mode respectively; $\hat{x}$ denotes the position operator
conjugate to the momentum operator $\hat{p}$; $|1>$ and
$|2>$ are the ground and excited states within the atom. According to
Sleator and Wilkens [16], one write the Hamiltonian for the atom-cavity
system with spatial motion
\begin{equation}
\hat{H}=\frac{\hat{p}^2}{2M}-\frac{\Delta}{2} (|1><1| - |2><2|) +
\omega a^{\dagger}a
+g(a^{\dagger} e^{-ikx} + ae^{ikx})(|1><2|+|2><1|)
\end{equation}
where $g$ is the atom-cavity coupling constant depending
on the mode-volume and the atomic dipole matrix elements. For simplicity,
we only consider the
effect of the spatial motion  of lower orders caused by the
long-period cavity field with small $k$.

As pointed out in refs.[7], the rotation-wave approximation is only
adequate to analyse the case of Jaynes-Cummings atom [19, 20]
close to resonance
and weak coupling, but the control of atomic tunneling requires the
case far away from resonance with a proper coupling. Thus, it is necessary
to develop an adaptable approximation method, which can work well in the
present situation. Fortunately, the generalized BO approximation
developed about four years ago for
the induced gauge structure and Berry's phase
[9, 10] can be extended here as a
systematically-analytical method to deal with the problem of the control for
atomic tunneling. The present  approach also recovers the adiabatic variational
principle used in ref.[7] as its lowest order approximate result.

By invoking an unitary transformation similar to that in ref.[6]
\begin{equation}
\hat{W}(x)=\exp{(-ikx a^\dagger a)}
\end{equation}
one obtains an approximate effective Hamiltonian $H_e=W^\dagger H W:$
\begin{equation}
\hat{H}_e = \frac{\hat{p}^2}{2M}+\frac{\Delta}{2}
(|2><2|-|1><1|)+\Omega(\hat p)
a^{\dagger} a +g(a+a^{\dagger})(|1><2|+|2><1|)
\end{equation}
with momentum-dependent  frequency
\begin{equation}
\hat{\Omega}=\Omega(\hat{p})=\omega - \frac{k\hat{p}}{M}
\end{equation}
Here, the effective frequency was modified by the Doppler shift
$\frac{k\hat{p}}{M}$ and the nonlinear term $k^2 (a^{\dagger} a)^2$ appearing
as the Kerr-like interaction has been neglected for
the consideration of large-period cavity field.
\vspace{1cm}

{\bf 3.Generalized BO approximation }
\vspace{0.5cm}

To describe the tunneling and localization of  the symmetric and antisymmetric
superpositions  of two atomic eigenstates
$$ |\pm>=\frac{1}{\sqrt{2}}(|1>\pm|2>)$$
we make  an ansatz
\begin{equation}
|\psi>=\phi_+ |+> + \phi_- |->
\end{equation}
for the  eigenstate of the atom-cavity system
by  drawing  an  analogy to the original BO approximate expansion [9,10].
Here, the vector-valued coefficients $\phi_{\pm}$ depend on both Fock space
of the cavity field and the spatial variable of the atomic mass-center.
They are imagined as the collective degree of freedom in the original
BO approximation. Substituting $|\psi>$ into the eigenvalue equation
$\hat{H}_e |\psi>=E|\psi>,$ one can obtain an operator-valued matrix equation
\begin{equation}
H\Phi +V\Phi = E\Phi
\end{equation}
with the definitions
\begin{equation}
H=\left( \matrix{\hat{H}_+ & 0\cr 0&\hat{H}_-\cr}\right ),V=\frac{1}{2}\Delta
\left( \matrix{0&1\cr 1&0\cr}\right), \Phi=\left(\matrix{\phi_+ \cr \phi_-
\cr}\right )
\end{equation}
where
\begin{equation}
\hat{H}_\pm=\frac{\hat{p}^2}{2M} + \frac{1}{2}\Omega(\hat{p}) a^{\dagger}a
\pm g(a+a^{\dagger})
\end{equation}
As proved in the generalized approach of BO approximation
[9,10], the usual stationary
perturbation theory associated with the above
representations (7-9) can result in the BO approximate solutions
for the coefficient vector $\Phi$ to any order of the  perturbation $V$.
For instance, the second order solutions are constructed as
\begin{equation}
\Phi_{\gamma}^{[1]} (n)=\sum_{n\ne m}
\frac{<\Phi_{\beta}^{[0]}(m,p)|\Phi_{\gamma}^{[0]}(n,p)>}
{2(E_n -E_m)} \Delta \Phi_{\beta}^{[0]} (m,p) ,\gamma\neq\beta=\pm
\end{equation}
from the first order ones
\begin{equation}
\Phi_{+}^{[0]} (n,p)=\left(\matrix{|\eta_+ (n)>\cr 0\cr}\right)\otimes|p>,
\Phi_{-}^{[0]} (n,p)=\left(\matrix{0\cr|\eta_- (n)>\cr}\right)\otimes |p>,
\end{equation}
with eigenvalues
\begin{equation}
E_{\pm,n} =E_n=\frac{p^2}{2M} + (n-\alpha^2)\Omega(p) ,
\alpha=\frac{g}{\Omega(p)}
\end{equation}
Here, $|\eta_\pm (n)>$ are determined by the Hamiltonian (9) as
the solutions
\begin{equation}
|\eta_\pm(n)>=|\eta_\pm (n,\alpha)=D(\mp \alpha)|n>.
\end{equation}
of  the eigen-equations
$$\hat{H}_\pm |\eta_\pm (n)>\otimes|p>=E_n |\eta_\pm (n)>\otimes|p>; $$
and
\begin{equation}
D(z)=exp[za^{\dagger}-z^* a]
\end{equation}
is the displace-operator of coherent state $|z>=D(z)|0>$; $ |p>$ is
a momentum eigenstate such that $\hat{p} |p>=p|p>$; $|n>$ is the Fock state.
It has to be pointed out that the stationary states $|\eta_\pm(n)>$
were even obtained with the adiabatic variational principle in ref.[7],
However, all the results in ref.[7] are only of the the special case for the
general ones in present studies, which, in fact, are of
first order in comparison with that in present studies.

With the help of straightforward substitutions of eq.(11) into eq.(10),
it is not difficult to calculate the coefficient $\phi_\pm$ to be of second
order
\begin{equation}
|\phi_\pm ^{[1]}(n,p)>= [|\eta_\pm (n) >\otimes |\pm> + \sum_{m\ne n}
\Delta^{\pm}_{mn} (p) |\eta_\mp (m)>\otimes|\mp>]\otimes|p>
\end{equation}
where
\begin{eqnarray}
\Delta^{\pm}_{mn} (p)=\frac{\Delta}{2\Omega (p)} \frac{<\eta_\mp (m)|
\eta_\pm (n)>}{m-n} = \frac{\Delta\cdot F(m,n)}{2\Omega (p) (m-n)}\nonumber\\
\equiv
\frac{\Delta \sqrt{m!n!}}{2(\omega - \frac{k\hat{p}}{M})}
\sum_{l=0}^{Min(m,n)}
\frac{e^{-|\alpha|^2 }(\mp 2\alpha)^{m+n-2l}}
{(m-l)! (m+l)! (m-n)!}
\end{eqnarray}
Notice that $|\phi_{\pm} (n,p)>$ are degenerate for a fixed energy $E_n$.
In the view of BO approximation, the first term in r.h.s. of eq.(15)
describes the adiabatic process and leads to an approximately-
stationary evolution for the  atomic
localization states $|\pm>$, that is to say, the evolving state starting
from $|+>$ (or $|->$) only differs from this initial state in a phase
factor with an invariant norm. In this sense
, one need to neglect the second term in the r.h.s. of eq.(15) whose norm
proportional to $ |\Delta^{\pm}_{mn}|$ under the adiabatic conditions
\begin{equation}
|\Delta^{\pm}_{mn}| \sim \frac{\Delta}{|\omega - \frac{pk}{M}|}\ll 1.
\end{equation}
This is just the stationary condition for the evolution of $|\pm>$
In fact, the second term in the r.h.s. of eq.(15) represents the
non-adiabatic effects of evolution. It is quite interesting that  whether
 the states $|\pm>$ are stationary or not must depend on the initial
momentum  of atom in the cavity, exactly speaking,
the velocity and the direction
of spatial motion of atomic mass-center. Therefore, as shown in next section,
the spatial motion of the atomic mass-center must
exercise an observable effect- the Doppler effects on the tunneling and
localization of atomic state.
\vspace{1cm}

{\bf 4. Control of atomic tunneling at zero temperature}
\vspace{0.5cm}

Recently, Plata and Gomez Lorent showed that the existence of cavity
field under certain conditions may decrease the effective energy-difference
between the dressed states of $|1>$ and $|2>$ so that  they   approach
degeneracy in  presence of the  cavity field [7]. In the present
sense,  the dressed states of
$|1>$ and $|2>$ are  modified by the spatial variable of atom besides the
cavity  mode and then
can also approach degeneracy only for the suitable momentum state of the atom's
mass center. Therefore,
they  evolve according to
Schrodinger equation with the approximately-equal phases to realize
the localization for the tunneling between the dressed states
$|+>$ and  $|->$ in presence of certain quantized cavity.
During this process, the dressed localization states are approximately
stationary. Indeed, the discussion in the section 3 demonstrates that the
first order dressed states
$$|n,\alpha,\pm,p>= |\eta_{\pm} (n)>\otimes |\pm>\otimes |p>$$
like the  approximate eigenstates of first order for $H_e$ possess
the approximately-equal
energies for the effective Hamiltonian (4).

To analyse dynamics quantitatively
for the problems mentioned above ,
the solutions (15) are
transformed back to the original representation
\begin{eqnarray}
|\psi_\pm^{[1]} (n,p)>=\hat W(x) |\phi_\pm^{[1]}(n,p)>\nonumber\\
=|\psi_\pm ^{[0]}(n,p)>+ \sum_{m\ne n}
\Delta^{\pm}_{mn} (p) |\psi_\pm^{[0]}(n,p]>
\end{eqnarray}
where
\begin{equation}
|\psi_\pm^{[0]}(n,p)>=\hat W |n,\alpha,\pm,p>\nonumber\\
=|\eta_\pm(n,\alpha e^{ikx})>\otimes|\pm>\otimes|p+ n k>.
\end{equation}
The above expression manifests that, under the first order approximation, the
approximately-stationary states are $|\psi^{[0]}_\pm (n,p)>$.
Unlike those for the case without spatial motion effect, the above stationary
states are not only dressed by the cavity field, but also accompanied
with the momentum shifts for the different components to $|n>$. Imaging
$|\psi_\pm^{[0]}(n,p)>$
as the right and left localization states in one-dimensional double-well
potential, one can consider the tunneling problem between $|\pm>$
and $|\mp>$.

Let us now focus on the simplest case that the atom is initially in
the ``left" dressed
state $|\psi_+^{[0]} (n,p)>$, in which  the cavity is in the  displace Fock
state
$|\eta_+ (n,\alpha e^{ikx})>$ while the atom in the ``left" state with
the momentum shift $p+n k$. Under the second approximation, the
wavefunction of cavity-atom system at $t$ is
\begin{eqnarray}
|\Psi_n (t)>=\hat W(x) \{ \exp(-i\frac{\hat{p}^2 t}{2M} +
i\alpha^2 \Omega t)\nonumber
\\
\times [e^{-i n\omega t} |\phi^{[1]}_+(n,p)> - \sum_{m\ne n}
\Delta^{+}_{mn}(p)
e^{-im\omega t} |\phi^{[1]}_-(n,p)>] \} \nonumber\\
= \exp[-\frac{i(p+ k a^\dagger a)^2 t}{2M}+i\alpha^2 \Omega t]
[e^{-i n \omega t} |\psi^{[0]}_+(n,p)> \nonumber\\
+ \sum_{m\ne n} \Delta^{+}_{mn}(p+ k a^\dagger a)
[e^{-i n\omega t}-e^{-im\omega t}]
 |\psi_-^{[0]}(m,p)>]
\end{eqnarray}
which gives the probability of the transition from  $|\psi_{+}^{[0]}(n,p)>$
to
$|->$
\begin{eqnarray}
P_n=\sum_{m\neq n}4|\Delta^{+}_{mn}|^2
\sin^2 [\frac{1}{2} (m-n)\omega t] \nonumber\\
=\sum_{m\neq n} \frac{\Delta^2 F(m,n)^2}
{(\omega - \frac{kp}{M})^2(m-n)^2}\sin^2 [\frac{1}{2} (m-n)\omega t].
\end{eqnarray}

Obviously, if one ignores  the second term proportional to
$\Delta_{m,n}^{\pm}$ in eq.(20), the atomic state $|+>$
is approximately stationary for
$$\sum_{n,p}|<\psi_-^{[0]} n,p|\Psi_m(t)>|^2\sim 0$$
It is also observed that the tunneling rate from a dressed state of $|+>$ to
$|->$
can be controlled by using the cavity field with suitable
frequency $\omega$ and preparing the atom with a proper initial
momentum. If the atom moves along the
direction opposite to the wave vector $\vec{k}$ in high frequency
cavity field, the tunneling rate (21) tends to be very small
and then the atomic dressed state
$|\psi_{\pm}^{[0]}(p)>$ tends to be
well localized in $|+>$. Preparing different initial state, e.g., the lower
frequency cavity and the atomic momentum along $\vec{k}$, the
localization will be broken and the tunneling process will be enhanced.

To complete a dynamical description of the tunneling control, one must also
specify the different initial conditions for the problem. Since various
initial conditions
for the single-mode field have become experimentally realizable,
it is useful to
obtain the different formulas of the corresponding tunneling rates. For
a general pure-state distribution of the cavity field
\begin{equation}
|c>=\sum C_n |\eta_+ (n)>,
\end{equation}
the tunneling rate is a superposition of those  oscillations with
different frequencies $(m-n)\omega$
\begin{equation}
P=\sum_n\sum_{m\ne n}|C_n|^2 \frac{\Delta^2\cdot
F(m,n)^2}{(\omega-\frac{kp}{M})^2(m-n)^2}
\sin^2[\frac{\omega (m-n)}{2}]
\end{equation}
When the coherent state $|Z>$ is taken as the initial state for the cavity,
the distributions $C_n$ are specified as
\begin{equation}
C_n (z) = <\eta_{+} (n)|z>=e^{-\frac{1}{2}\alpha(z-z^*)} e^{-\frac{1}{2}
|\alpha +
z|^2} \frac{(\alpha + z)^n}{\sqrt{n!}}
\end{equation}
where $ |z|^2 $ denotes the mean photon number. If the initial state has
definite
photon number, that is $|m>$, then one has the specific distribution
\begin{eqnarray}
C_n=C_{n;m}(\alpha)=<\eta_{+} (n)|m>=<n|D(\alpha)|m>\nonumber\\
=\sum_{l=0}^{Min(n,m)} \frac{\sqrt{n! m!}}{(n-l)!(m-l)!} e^{-\frac{1}{2}
|\alpha|^2}
\alpha^{n+m-2l};
\end{eqnarray}
\vspace{1cm}

{\bf 5. The cases with finite temperature }

In this section, we turn to discuss the influence of the temperature $ T $ of
the
cavity on the dynamics of tunneling control. In this sense, the cavity is
supposed in thermal equilibrium and then the corresponding mixed state
described by Bose-Einstein photon number distribution
\begin{equation}
\rho_c (0)=\sum_{n=0}^{\infty} \frac{1}{\Omega} e^{-n\beta \omega} |n><n|
\end{equation}
where
$$\beta=\frac{1}{kT},~~~\Omega = (1-e^{-\beta \omega})^{-1}.$$
Expressing the Fock states $|n>$ in terms of the displaced Fock states
$|\eta_{+} (n,\alpha e^{-ikt})>$,
one have an initial state for the atom-cavity system
\begin{equation}
\rho (0)=\sum_{n=0}^{\infty}\sum_{m,m'\ne n} \frac{1}{\Omega}
e^{-n\beta \omega}
{C'}_{m,n}{C'}_{m,'n}^{*} |\psi^{[0]}_+(m,p-mk)><\psi^{[0]}_+
(m',p-m'k\hbar)|
\end{equation}
where ${C'}_{m,n}=C_{m,n}(\alpha e^{ikx})$ is defined by eq.(25) and the
initial state of atomic mass-center
is chosen
as $|+>\otimes|p> $. Then, one can write the density matrix for the atom-cavity
system at time $t$.
\begin{equation}
\rho (t)=U(t)\rho (0) U(t)^\dagger
= \sum^{\infty}_{n=0}\sum_{m',m\ne n} \frac{1}{\Omega} e^{-\beta n\omega}
{C'}_{mn}
{C'}^*_{m'n} |{\Psi'}_m (t)><{\Psi'}_{m}' (t)|
\end{equation}
where
$$|{\Psi'}_m (t)>=|\Psi_m(t)>|_{p\rightarrow p-mk}$$
is given by eq.(20) and $U(t)$ is the evolution operator

Then, the tunneling rate at temperature $T$ is obtained from eq.(28) as the
probability of finding atom in the state $|->$:
\begin{equation}
P(T,t)=Tr(|-><-| \rho (t))\nonumber\\
=\sum \frac{1}{\Omega} e^{-\beta n\omega} P_n (t)
\end{equation}
where
\begin{eqnarray}
P_n (t)=Tr (<-|U(t)|n,+,p><n,+,p|U(t)|->)\nonumber \\
=\sum_{m,m'} {C'}_{m,n}{C'}_{m',n}^*Tr(<-|{\Psi'}_{m}(t)><{\Psi'}_{m'}
(t)|->\nonumber\\
=\sum_{l=0} |\sum_{m\ne l} C_{m,n}'\Delta_{lk}^{+}(p-mk) [e^{-im\omega t }
-e^{-il\omega t}]|^2
\end{eqnarray}
is the transition probability for n'th channel switched on by the existence of
the thermal cavity field where
$$|n,+,p>=|n>\otimes|+>\otimes |p>$$

Obviously, if we have many identical atoms in the cavity with single-mode
radiation in thermal equilibrium with the wells at temperature $T$, the
tunneling rate will increase as $T $ becomes higher and then
the thermal perturbation must enhances the tunneling. Conversely, at the lower
temperature, the tunneling rate is suppressed and then the localization of
state
$|\phi_{+} (p)>$ is easily realized. Therefore, the experiment to control
tunneling
and localization should be well carried out at lower temperature. This
is a trivial but very useful observation.
\vspace{2cm}

{\bf Acknowledgement}

The author is much grateful to Professor C.N.Yang for many encouragements and
drawing his attention to the new field in modern atomic physics-
the cavity-quantum electrodynamics.
This work is partially supported by the NSF of China and the Fok Yin-Tung
Education Foundation.

\end{document}